\documentclass[12pt]{article}
  \usepackage{amsfonts}
  \usepackage{amsmath}
\usepackage{amssymb}
\usepackage{amscd}
\usepackage[dvips]{graphicx}

\begin{document}

~~
\bigskip
\bigskip
\begin{center}
{\Large {\bf{{{Saturation of uncertainty relations for
twisted acceleration-enlarged Newton-Hooke space-times}}}}}
\end{center}
\bigskip
\bigskip
\bigskip
\begin{center}
{{\large ${\rm {Marcin\;Daszkiewicz}}$}}
\end{center}
\bigskip
\begin{center}
\bigskip

{${\rm{Institute\; of\; Theoretical\; Physics}}$}

{${\rm{ University\; of\; Wroclaw\; pl.\; Maxa\; Borna\; 9,\;
50-206\; Wroclaw,\; Poland}}$}

{ ${\rm{ e-mail:\; marcin@ift.uni.wroc.pl}}$}
\end{center}
\bigskip
\bigskip
\bigskip
\bigskip
\bigskip
\bigskip
\bigskip
\bigskip
\bigskip
\begin{abstract}
  Using Fock representation we construct  states saturating uncertainty relations for twist-deformed
acceleration-enlarged Newton-Hooke space-times.
\end{abstract}
\bigskip
\bigskip
\bigskip
\bigskip
\eject

\section{Introduction}

In the last time, there appeared a lot of papers dealing with 
classical and quantum  mechanics (see e.g. \cite{mech}-\cite{qmnext})
as well as with field theoretical models (see e.g. \cite{field}), in
which  the quantum space-time  plays a crucial role.  
The idea to use noncommutative coordinates is quite old - it goes back to Heisenberg
and was firstly formalized by Snyder in   \cite{snyder}.  Recently, however, there were found new formal
arguments based mainly on Quantum Gravity \cite{grav1} 
and String Theory models \cite{string1}, 
indicating that space-time at Planck scale  should be
noncommutative, i.e. it should  have a quantum nature. Besides, 
 the main reason for such considerations follows from the
suggestion  that relativistic space-time symmetries should be
modified (deformed) at Planck scale, while  the classical Poincare
invariance still remains valid at
larger distances \cite{1a}, \cite{1anext}.

Currently, it is well known, that in accordance with the
Hopf-algebraic classification of all deformations of relativistic
and nonrelativistic symmetries, one can distinguish three 
types of quantum spaces \cite{class1}, \cite{class2} (for details see also \cite{conservative}):\\
\\
{ \bf 1)} Canonical ($\theta^{\mu\nu}$-deformed) type of quantum space \cite{oeckl}-\cite{dasz1}
\begin{equation}
[\;{ x}_{\mu},{ x}_{\nu}\;] = i\theta_{\mu\nu}\;, \label{noncomm}
\end{equation}
\\
{ \bf 2)} Lie-algebraic modification of classical space-time \cite{dasz1}-\cite{lie1}
\begin{equation}
[\;{ x}_{\mu},{ x}_{\nu}\;] = i\theta_{\mu\nu}^{\rho}{ x}_{\rho}\;,
\label{noncomm1}
\end{equation}
and\\
\\
{ \bf 3)} Quadratic deformation of Minkowski and Galilei  spaces \cite{dasz1}, \cite{lie1}-\cite{paolo}
\begin{equation}
[\;{ x}_{\mu},{ x}_{\nu}\;] = i\theta_{\mu\nu}^{\rho\tau}{
x}_{\rho}{ x}_{\tau}\;, \label{noncomm2}
\end{equation}
with coefficients $\theta_{\mu\nu}$, $\theta_{\mu\nu}^{\rho}$ and  $\theta_{\mu\nu}^{\rho\tau}$ being constants.\\
\\
Besides, it has been demonstrated in \cite{nh}, that in the case of
so-called acceleration-enlarged Newton-Hooke Hopf algebras
$\,{\mathcal U}_0(\widehat{ NH}_{\pm})$ the twist deformation
provides the new  space-time noncommutativity of the
form\footnote{$x_0 = ct$.},\footnote{ The discussed space-times have been  defined as the quantum
representation spaces, so-called Hopf modules (see \cite{oeckl}, \cite{chi}, 
\cite{bloch}, \cite{wess}), for quantum acceleration-enlarged
Newton-Hooke Hopf algebras.},\footnote{The twisted (usual) Newton-Hooke quantum
 space-times have been provided in  \cite{nh1}.}
\begin{equation}
{ \bf 4)}\;\;\;\;\;\;\;\;\;[\;t,{ x}_{i}\;] = 0\;\;\;,\;\;\; [\;{ x}_{i},{ x}_{j}\;] = 
if_{\pm}\left(\frac{t}{\tau}\right)\theta_{ij}(x)
\;, \label{nhspace}
\end{equation}
with time-dependent  functions
$$f_+\left(\frac{t}{\tau}\right) =
f\left(\sinh\left(\frac{t}{\tau}\right),\cosh\left(\frac{t}{\tau}\right)\right)\;\;\;,\;\;\;
f_-\left(\frac{t}{\tau}\right) =
f\left(\sin\left(\frac{t}{\tau}\right),\cos\left(\frac{t}{\tau}\right)\right)\;,$$
$\theta_{ij}(x) \sim \theta_{ij} = {\rm const}$ or
$\theta_{ij}(x) \sim \theta_{ij}^{k}x_k$ and  $\tau$ denoting the time scale parameter 
 -  the cosmological constant. It should be also noted that different relations  between all mentioned above quantum spaces ({\bf 1)}, { \bf 2)}, { \bf 3)} 
and { \bf 4)}) have been summarized in paper \cite{conservative}.

From historical point of view, the studies  on
so-called coherent states were started by Schr\"{o}dinger,
who minimalized uncertainty relations for position and momenta operator in the case of harmonic oscillator
model \cite{sch}. The result of these investigations has been applied in the 60's by Glauber to provide a complete quantum-theoretical
description of coherence for electromagnetic free field \cite{glaubner}. It was a pioneer work in quantum optic theory
describing  phenomena associated with such processes as laser light emission or laser interferometry  \cite{ligo}.  Recently, in articles \cite{23} and \cite{23a}, the above-mentioned results  have been extended to the case of canonically deformed
space-time (\ref{noncomm}). Particulary, it has been constructed the proper Fock space of quantum states and
the deformed coherent wave functions.

In this article, following \cite{23} and \cite{23a}, we find coherent states for twisted acceleration-enlarged Newton-Hooke space-times (\ref{spaces}), i.e.
we provide states which saturate the deformed uncertainty relations (\ref{nnh2a})-(\ref{nnh2c}). In first section we recall
basic facts associated with saturation of Heisenberg relations for commutative space-time. Section 2  concerns the saturation of twist-deformed uncertainty relations (\ref{nnh2a})-(\ref{nnh2c}) - it contains the construction of Fock space and twisted coherent states. The final remarks are discussed in the last section.

\section{Saturation of uncertainty relations and coherent states in commutative space-time}

\subsection{General prescription}

Let us start with  general algorithm  for saturation of uncertainty principles described
 in \cite{24} and \cite{24a}. Hence, it is well-known that for arbitrary
two observables $\hat{a}$, $\hat{b}$ such that
\begin{equation}
\label{ww1}
[\;\hat{a},\hat{b}\;]=i\hat{c}\;,
\end{equation}
one can derive the following (so-called generalized Heisenberg principle) inequality
\begin{equation}
\label{ww2}
(\Delta \hat{a})_\psi\cdot(\Delta \hat{b})_\psi\geq\frac{1}{2}|< \hat{c}>_\psi|\;,
\end{equation}
where $|\psi >$ denotes quantum state normalized to unity and
\begin{equation}
\label{ww3}
(\Delta \hat{o})_\psi=\sqrt{<\psi|(\hat{o}\;-<\hat{o}>_\psi \mathbb{I})^2|\psi>}\;\;;\;\;\hat{o} =
\hat{a}, \hat{b}\;.
\end{equation}
The Heisenberg relation (\ref{ww2}) is saturated  when the following condition is
satisfied $(\xi \in \bf{R})$
\begin{align}
\label{ww4}
(\hat{a}\;-<\hat{a}>_\psi \mathbb{I})|\psi>=-i\xi(\hat{b}\;-<\hat{b}>_\psi \mathbb{I})
|\psi>\;,
\end{align}
Further, by acting with $\hat{a}\;-<\hat{a}>_\psi \mathbb{I}$ on both sides of equation (\ref{ww4}), using formula
(\ref{ww1}) and again (\ref{ww4}),  one can rewrite the above condition as follows 
\begin{equation}
\label{ww5}
(\hat{a}\;-<\hat{a}>_\psi \mathbb{I})^2|\psi>=-\xi^2(\hat{b} \;-<\hat{b}>_\psi \mathbb{I})^2
|\psi>+\xi\hat{c}|\psi>\;,
\end{equation}
or, equivalently, on multiplying by $|\psi>$ from the left as
\begin{equation}
\label{ww6}
(\Delta \hat{a})^2_\psi+\xi^2(\Delta \hat{b})^2_\psi=\xi < \hat{c}>_\psi\;.
\end{equation}
It is easy to check that the relation (\ref{ww6}) together with the saturated form of
 Heisenberg principle (\ref{ww2}) gives
\begin{align}
\label{ww7}
(\Delta \hat{a})^2_\psi=\frac{\xi}{2}< \hat{c} >_\psi\;\;\;,\;\;\;
(\Delta \hat{b})^2_\psi=\frac{1}{2\xi}< \hat{c} >_\psi\;,
\end{align}
which explains the meaning of $\xi$.

\subsection{The standard Heisenberg relation case}

Let us now apply the above scheme to the standard Heisenberg relation
\begin{equation}
\label{ww8}
[\;\hat{x},\hat{p}\;]=i\hbar\;,
\end{equation}
which yields  inequality
\begin{equation}
\label{ww9}
\Delta \hat{x}\cdot\Delta \hat{p}\geq\frac{\hbar}{2}\;.
\end{equation}
In accordance with the formula (\ref{ww4}) one can observe that  the uncertainty relation
(\ref{ww9}) is saturated iff
\begin{align}
\label{ww10}
(\hat{x}-\alpha \mathbb{I})|\psi>=-i\xi (\hat{p}-\beta\mathbb{I})|\psi>\;,
\end{align}
where $\alpha=<\hat{x} >_\psi$ and $\beta=< \hat{p}>_\psi$.  Next, we
 define in a standard way the  creation/annihilation operators\footnote{We use $\omega=m=1$ units.}
\begin{equation}
\label{ww11}
\begin{split}
a\equiv\frac{1}{\sqrt{2\hbar}}(\hat{x}+i\hat{p})\;\;\;,\;\;\;
a^\dagger\equiv\frac{1}{\sqrt{2\hbar}}(\hat{x}-i\hat{p})\;,
\end{split}
\end{equation}
satisfying
\begin{equation}
\label{ww11a}
[\;a,a^{\dagger}\;]=1\;,
\end{equation}
and then, the Hilbert space of states is spanned by the vectors
\begin{equation}
\label{ww12}
|n>=\frac{1}{\sqrt{n!}}(a^\dagger)^n|0>\;.
\end{equation}
Firstly,
in order to find the general solution of equation (\ref{ww10}) one  should notice that $\xi$ is bigger than zero\footnote{In fact, parameter $\xi$ is different than zero because operator $\hat{x}-\alpha \mathbb{I}$ cannot have normalized eigenvectors (operators commuting to $c$-number have no normalized eigenvectors in their common invariant domain). Consequently, for $\xi \neq 0$ equation (\ref{ww7}) gives $\xi$ bigger than zero.}. Further, we consider  $\xi=1$ and observe that in such a case the equation (\ref{ww10}) can be rewritten as follows
\begin{align}
\label{ww13}
a|\psi>=z|\psi>\;\;\;{\rm with}\;\;\;z=\frac{\alpha+i\beta}{\sqrt{2\hbar}}\;.
\end{align}
The solutions of (\ref{ww13}), i.e. the eigenstates of  annihilation operator $a$ are called coherent states and, particulary,
 the vacuum vector is  coherent state corresponding to the eigenvalue  $z$ equal zero.
In order to find remaining solutions of (\ref{ww13}) one defines, for any complex value of $z$, the unitary operators
\begin{equation}
\label{ww14}
U(z)\equiv {\rm e}^{za^\dagger-\bar{z}a}={\rm e}^{-\frac{1}{2}|z|^2}{\rm e}^{za^\dagger}{\rm e}^{-\bar{z}a}\;.
\end{equation}
Next, one  easily  check that
\begin{equation}
\label{ww15}
U^\dagger(z)aU(z)=a+z\cdot \mathbb{I}\;,
\end{equation}
what means that any coherent state for $\xi = 1$ is given by
\begin{equation}
\label{ww16}
|z>\equiv U(z)|0>={\rm e}^{-\frac{1}{2}|z|^2}{\rm e}^{za^{\dagger}}|0>=
{\rm e}^{-\frac{1}{2}|z|^2}\sum_{n=0}^\infty \frac{z^n}{\sqrt{n!}}|n>\;.
\end{equation}
Let us now turn to the case $\xi\neq 1$ for which formula (\ref{ww10}) can be written as
\begin{equation}
\label{ww17}
a_\xi|\psi>=z|\psi>\;,
\end{equation}
with
\begin{equation}
\label{ww18}
\begin{split}
a_\xi&=\frac{1}{\sqrt{2\hbar}}\left(\frac{\hat{x}}{\sqrt{\xi}}+i\sqrt{\xi}\hat{p}\right)\;\;\;,\;\;\;
a_\xi^\dagger=\frac{1}{\sqrt{2\hbar}}\left(\frac{\hat{x}}{\sqrt{\xi}}-i\sqrt{\xi}\hat{p}\right)\;,
\end{split}
\end{equation}
and
\begin{equation}
\label{ww18a}
z=\frac{1}{\sqrt{2\hbar}}\left(\frac{\alpha}{\sqrt{\xi}}+i\beta\sqrt{\xi}\right)\;.
\end{equation}
It is also easy to verify that
\begin{equation}
\label{ww18b}
[\;a_\xi, a_\xi^\dagger\;]=1 \;,
\end{equation}
and that for $\xi =1$ we have
\begin{equation}
\label{ww18c}
a_{\xi=1}=a\;.
\end{equation}
Solutions of equation (\ref{ww17}) can be find
with use of $\xi$-creation/annihilation $a_\xi^\dagger$/$a_\xi$ operators and $\xi$-vacuum state $|0>_\xi$. However,
all representations of Fock algebra are unitarily equivalent and, indeed, one can check that
\begin{equation}
\label{ww20}
V(\xi)aV^\dagger(\xi) =a_\xi\;\;\;,\;\;\;
V(\xi)a^\dagger V^\dagger(\xi)=a_\xi^\dagger\;,
\end{equation}
for the unitary operator  $V(\xi)$ defined by
\begin{equation}
\label{ww19}
V(\xi)={\rm e}^{-\frac{1}{4}\ln\xi(a^2-(a^\dagger)^2)}\;,
\end{equation}
Consequently, the solution of  equation (\ref{ww10}) can be  written as
\begin{equation}
\label{ww21}
|z,\xi>=V(\xi)U(z)|0> = {\rm e}^{-\frac{1}{2}|z|^2}{\rm e}^{-\frac{1}{4}\ln\xi(a^2-(a^\dagger)^2)}{\rm e}^{za^{\dagger}}|0>\;,
\end{equation}
with complex parameter $z$  related to the mean values of $\hat{x}$ and $\hat{p}$ operators, and  $\xi$ describing their
dispersions (see formulas (\ref{ww7}) and (\ref{ww13}) respectively)
\begin{equation}
\label{ww22}
(\Delta \hat{x})^2=\frac{\xi\hbar}{2}\;\;\;,\;\;\;
(\Delta \hat{p})^2=\frac{\hbar}{2\xi}\;.
\end{equation}

\section{Coherent states for twist-deformed acceleration-enlarged Newton-Hooke space-times}

In this section we turn to the twisted acceleration-enlarged Newton-Hooke space-times  equipped with classical time and quantum spatial directions, i.e. we consider  spaces of the form
\begin{equation}
[\;t,\bar{x}_{i}\;] = 0\;\;\;,\;\;\; [\;\bar{x}_{1},\bar{x}_{2}\;] =
if({t})\;\;;\;\;i=1,2
\;, \label{spaces}
\end{equation}
with positive defined function $f({t})$ given by\footnote{$\kappa_a>0$.}
\begin{eqnarray}
f({t})&=&f_{\kappa_1}({t}) =
f_{\pm,\kappa_1}\left(\frac{t}{\tau}\right) = \kappa_1\,C_{\pm}^2
\left(\frac{t}{\tau}\right)\;, \label{w2}\\
f({t})&=&f_{\kappa_2}({t}) =
f_{\pm,\kappa_2}\left(\frac{t}{\tau}\right) =\kappa_2\tau^2\,
S_{\pm}^2 \left(\frac{t}{\tau}\right) \;, \label{w4}\\
f({t})&=&f_{\kappa_3}({t}) =
 f_{\pm,\kappa_3}\left(\frac{t}{\tau}\right) = 4\kappa_3
 \tau^4\left(C_{\pm}\left(\frac{t}{\tau}\right)
-1\right)^2 \;; \label{w7}
\end{eqnarray}
$$C_{+/-} \left(\frac{t}{\tau}\right) = \cosh/\cos \left(\frac{t}{\tau}\right)\;\;\;{\rm and}\;\;\;
S_{+/-} \left(\frac{t}{\tau}\right) = \sinh/\sin
\left(\frac{t}{\tau}\right) \;.$$
As it was already mentioned in Introduction, in $\tau \to \infty$ limit, the above quantum spaces reproduce the canonical (\ref{noncomm}),
 quadratic (\ref{noncomm2})  and quartic  type of
space-time noncommutativity, with\footnote{Space-times (\ref{nw2}), (\ref{nw4}) correspond to the
twisted Galilei Hopf algebras provided in \cite{dasz1}, while the quantum space (\ref{nw7}) is
associated with acceleration-enlarged Galilei Hopf structure \cite{nh}.} 
\begin{eqnarray}
f_{\kappa_1}({t}) &=& \kappa_1\;,\label{nw2}\\
f_{\kappa_2}({t}) &=& \kappa_2\,t^2\;,\label{nw4}\\
f_{\kappa_3}({t}) &=& \kappa_3\,t^4\;.  \label{nw7}
\end{eqnarray}
Of course, for all  parameters $\kappa_a$ running to zero the above deformations disappear.

The above spaces can be extended to the whole algebra of position and momentum operators as follows
\begin{eqnarray}
[\;\bar{ x}_{1},\bar{ x}_{2}\;] = if_{\kappa_a}({t})\;\;\;,\;\;\; [\;\bar{ p}_{i},\bar{ p}_{j}\;] =0\;\;\;,\;\;\;
[\;\bar{ x}_{i},\bar{ p}_{j}\;] = {i\hbar}\delta_{ij}\;\;;\;\;i,j=1,2\;, \label{phasespaces}
\end{eqnarray}
and then, the corresponding uncertainty relations take the form
\begin{eqnarray}
\Delta \bar{x}_1\Delta \bar{x}_2&\geq&\frac{f_{\kappa_a}({t})}{2}\;,\label{nnh2a}\\
\Delta \bar{x}_1\Delta \bar{p}_1&\geq&\frac{\hbar}{2}\;,\label{nnh2b}\\
\Delta \bar{x}_2\Delta \bar{p}_2&\geq&\frac{\hbar}{2}\;.\label{nnh2c}
\end{eqnarray}
In next two subsections we construct the quantum-mechanical states saturating the deformed Heisenberg
principles (\ref{nnh2a})-(\ref{nnh2c}). Partially, we use algorithm described in pervious section and the results of articles
\cite{23} and \cite{23a}.

\subsection{Oscillator representations}

In order to find the coherent states associated with twisted commutation relations (\ref{phasespaces}) we provide  their oscillator (irreducible) representations. First of all, we observe that position and
momentum operators $\bar{x}_i$ and $\bar{p}_i$ can be written in terms of canonical ones ($\hat{x}_i$, $\hat{p}_i$)
as follows
\begin{equation}
\label{nnh3}
\bar{x}_i \equiv \hat{x}_i-\frac{f_{\kappa_a}(t)}{2\hbar}\epsilon_{ij}\hat{p}_j\;\;\;,\;\;\;
\bar{p}_i\equiv\hat{p}_i\;,
\end{equation}
with $\epsilon_{12} = -\epsilon_{21} =1$ and $\epsilon_{11}=\epsilon_{22}=0$. Then, it seems sensible to introduce
the following definition of  creation/anihilation operators
\begin{eqnarray}
a_i(t)&\equiv& \frac{1}{\sqrt{2\hbar}}\left[\bar{x}_i+\left(i\delta_{ij}+\frac{f_{\kappa_a}(t)}{2\hbar}
\epsilon_{ij}\right)\bar{p}_j\right]\;,\label{nnh4a}\\
a_i^\dagger (t)&\equiv& \frac{1}{\sqrt{2\hbar}}\left[\bar{x}_i+\left(-i\delta_{ij}+\frac{f_{\kappa_a}(t)}{2\hbar}
\epsilon_{ij}\right)\bar{p}_j\right]\;,\label{nnh4b}
\end{eqnarray}
which satisfy
\begin{equation}
\label{nnh5}
[\;a_i(t),a_j^\dagger(t)\;]=\delta_{ij}\;.
\end{equation}
In such a way we arrive at   Fock space spanned by the orthonormal vectors of the form
\begin{equation}
\label{nnh6}
|n_1,n_2,t>=\frac{1}{\sqrt{n_1!}}\frac{1}{\sqrt{n_2!}}(a_1^\dagger(t))^{n_1}
(a_2^\dagger(t))^{n_2}|0>\;.
\end{equation}
For later convenience we also provide the modified operators
\begin{equation}
\label{nnh8}
a_{\pm}(t)\equiv\frac{1}{\sqrt{2}}(a_1(t)\mp ia_2(t))\;\;\;,\;\;\;
a_{\pm}^\dagger(t)\equiv\frac{1}{\sqrt{2}}(a_1^\dagger(t)\pm ia_2^\dagger(t))\;,
\end{equation}
leading to the following (new) basis
\begin{equation}
\label{w9}
|n_{+},n_{-},t>=\frac{1}{\sqrt{n_{+}!}}\frac{1}{\sqrt{n_{-}!}}(a_{+}^\dagger(t))^{n_{+}}
(a_{-}^\dagger(t))^{n_{-}}|0>\;.
\end{equation}

\subsection{Saturating of uncertainty relations}

Let us construct  all states which  saturate the uncertainty relations (\ref{nnh2a}). In this aim, in accordance
with algorithm proposed in \cite{23} and \cite{23a}, we define the following set of independent
creation/anihilation operators
\begin{equation}
\label{nnh11}
\begin{split}
b(t)&\equiv\sqrt{\frac{\hbar}{2f_{\kappa_a}(t)}}\left[
\left(1+\frac{f_{\kappa_a}(t)}{2\hbar}\right)a_{-}+
\left(1-\frac{f_{\kappa_a}(t)}{2\hbar}\right)a_{+}^\dagger\right]\;,\\
b^\dagger(t)&\equiv\sqrt{\frac{\hbar}{2f_{\kappa_a}(t)}}\left[
\left(1+\frac{f_{\kappa_a}(t)}{2\hbar}\right)a_{-}^\dagger+
\left(1-\frac{f_{\kappa_a}(t)}{2\hbar}\right)a_{+}\right]\;,\\
c(t)&\equiv\sqrt{\frac{\hbar}{2f_{\kappa_a}(t)}}\left[
\left(1+\frac{f_{\kappa_a}(t)}{2\hbar}\right)a_{+}+
\left(1-\frac{f_{\kappa_a}(t)}{2\hbar}\right)a_{-}^\dagger\right]\;,\\
c^\dagger(t)&\equiv\sqrt{\frac{\hbar}{2f_{\kappa_a}(t)}}\left[
\left(1+\frac{f_{\kappa_a}(t)}{2\hbar}\right)a_{+}^\dagger+
\left(1-\frac{f_{\kappa_a}(t)}{2\hbar}\right)a_{-}\right]\;.
\end{split}
\end{equation}
Next, by straightforward calculations we get
\begin{equation}
\label{nnh12}
b(t)=\frac{1}{\sqrt{2f_{\kappa_a}(t)}}(\bar{x}_1+i\bar{x}_2)\;\;\;,\;\;\;
b^\dagger(t)=\frac{1}{\sqrt{2f_{\kappa_a}(t)}}(\bar{x}_1-i\bar{x}_2)\;,
\end{equation}
what means that both $b$-operators are spanned in a standard way by noncommutative positions
$\bar{x}_1$ and $\bar{x}_2$. Consequently, due to the commutation relations (\ref{spaces})\footnote{They
are the same as canonical commutation relations (\ref{ww8}) with operator $\hat{p}$
replaced by $\hat{x}_2$ and $\hbar$ replaced by function $f_{\kappa_a}(t)$.} one can applied the
standard scheme proposed in Section 2. 
Then, in accordance with formula (\ref{ww21})  we have
\begin{equation}
\label{nnh13}
|z,\xi,t>={\rm e}^{-\frac{1}{2}|z|^2}{\rm e}^{+\frac{1}{4}\ln\xi((b^\dagger(t))^2-b^2(t))}
{\rm e}^{zb^\dagger(t)}|0,t>_b\;,
\end{equation}
where symbol $|0,t>_b$ denotes the vacuum state for annihilator $b(t)$, i.e.
\begin{equation}
\label{nnh14}
b(t)|0,t>_b=0\;.
\end{equation}
It should be noted that the choice of state (\ref{nnh14}) is not unique, i.e. it may contain an arbitrary
number of $c^\dagger(t)$ excitation. Besides, one should observe that in accordance with formal arguments
proposed in \cite{23}, \cite{23a}, the representation given by $b(t)$, $b^\dagger(t)$, $c(t)$ and $c^\dagger(t)$
 operators is unitary equivalent to
that defined by $a_\pm(t)$ and $a_\pm^\dagger(t)$. The corresponding transformation is given by\footnote{It can be find
by analogy to the algorithm proposed in \cite{23} and \cite{23a} for the case of canonical deformation (\ref{noncomm}).}
\begin{eqnarray}
b(t)&=&T(t)\left[a_{-}(t)\right]T^\dagger(t)\;,\label{nnh15a}\\
b^\dagger(t)&=&T(t)\left[a^\dagger_{-}(t)\right]T^\dagger(t)\;,\label{nnh15b}\\
c(t)&=&T(t)\left[a_{+}(t)\right]T^\dagger(t)\;,\label{nnh15c}\\
c^\dagger(t)&=&T(t)\left[a^\dagger_{+}(t)\right]T^\dagger(t)\;,\label{nnh15d}
\end{eqnarray}
with
\begin{equation}
\label{nnh16}
T(t)={\rm e}^{\frac{1}{2}\ln\left(\frac{2\hbar}{f_{\kappa_a}(t)}\right)(a_{+}(t)a_{-}(t)-
a_{+}^\dagger(t) a_{-}^\dagger(t))}\;.
\end{equation}
Consequently, it means that the states   saturating (\ref{spaces}) are linear combinations
with respect to $n_{+}$  of the vectors\footnote{$z$ and $\xi$ are fixed.}
\begin{equation}
\label{nnh21}
|z,\xi,n_{+},t>={\rm e}^{-\frac{1}{2}|z|^2}T(t)
{\rm e}^{-\frac{1}{4}\ln\xi(a_{-}^2(t)-(a_{-}^\dagger(t))^2)}
{\rm e}^{za_{-}^\dagger(t)}|n_{+},0,t>\;.
\end{equation}

Let us now turn to the states saturating
\begin{equation}
\label{nnh22}
\Delta \bar{x}_1\Delta \bar{p}_1\geq \frac{\hbar}{2}\;.
\end{equation}
Firstly, as in the pervious case, we define the new creation/anihilation operators as follows
\begin{eqnarray}
d(t)&=&a_1(t)+\frac{if_{\kappa_a}(t)}{4\hbar}(a_2(t)-a_2^\dagger(t))\;,\label{nnh23a}\\
d^\dagger(t)&=&a^\dagger_1(t)+\frac{if_{\kappa_a}(t)}{4\hbar}(a_2(t)-a_2^\dagger(t))\;,  \label{nnh23b}\\
e(t)&=&a_2(t)+\frac{if_{\kappa_a}(t)}{4\hbar}(a_1(t)-a_1^\dagger(t))\;,\label{nnh23c}\\
e^\dagger (t)&=&a_2^\dagger(t)+\frac{if_{\kappa_a}(t)}{4\hbar}(a_1(t)-a_1^\dagger(t))\;,\label{nnh23d}
\end{eqnarray}
which in $d$-sector take the form
\begin{equation}
\label{nnh24}
d(t)=\frac{1}{\sqrt{2\hbar}}(\bar{x}_1+i\bar{p}_1)\;\;\;,\;\;\;d^\dagger(t)=\frac{1}{\sqrt{2\hbar}}
(\bar{x}_1-i\bar{p}_1)\;.
\end{equation}
Further, one can find unitary transformation connecting $d(t)$ and $e(t)$ operators with old ones.
It looks as follows
\begin{equation}
\label{nnh25}
d(t)=S(t)\left[a_1(t)\right]S^\dagger(t)\;\;\;,\;\;\;
e(t)=S(t)\left[a_2(t)\right]S^\dagger(t)\;,
\end{equation}
where
\begin{equation}
\label{nnh26}
S(t)={\rm e}^{\frac{if_{\kappa_a}(t)}{4\hbar}(a_1(t)-a_1^\dagger(t))(a_2(t)-a_2^\dagger(t))}\;.
\end{equation}
Consequently, the states saturating (\ref{nnh22}) can be written as linear combinations,
with respect to $n_2$ but with parameters $z$ and $\xi$ fixed, of the following vectors
\begin{equation}
\label{nnh27}
|z,\xi,n_2,t>={\rm e}^{-\frac{1}{2}|z|^2}S(t){\rm e}^{-\frac{1}{4}\ln\xi(a_1(t)^2-(a_1^\dagger(t))^2)}
{\rm e}^{za_1^\dagger(t)}|0,n_2,t>\;,
\end{equation}
It is easy to see, that the states saturating (\ref{nnh2c}) are obtained by exchanging index "1" to "2" and functions
$f_{\kappa_a}(t)$ to $-f_{\kappa_a}(t)$ in formula (\ref{nnh27}), i.e.
\begin{equation}
\label{nnh28}
|z,\xi,n_1,t>={\rm e}^{-\frac{1}{2}|z|^2}S^\dagger(t)
{\rm e}^{-\frac{1}{4}\ln\xi(a_2(t)^2-(a_2^\dagger(t))^2)}
{\rm e}^{za_2^\dagger(t)}|0,n_1,t>\;.
\end{equation}

\section{Final remarks}

In this article we construct  states saturating uncertainty relations for twisted
acceleration-enlarged Newton-Hooke space-times (\ref{nhspace}). Particulary, for very special choice of  considered quantum spaces we get the results obtained in \cite{23} and \cite{23a} for canonical deformation (\ref{noncomm}).

It should be noted that presented investigation has been performed in the case of quite simple deformation with two
spatial directions commuting to  function of classical time. However, the mentioned studies can be extended in
nontrivial way to much more complicated space-time models, such as -  Lie-algebraic or quadratic type of noncommutativity
with two spatial directions commuting to space. Besides, one should better understand the obtained results in context of
(for example) wave-gravitational interferometry processes, for which saturating states play a prominent role \cite{ligo}. The works in these
directions already started and are in progress.

\section*{Acknowledgments}
The author would like to thank J. Lukierski and A. Frydryszak
for valuable discussions.


\begin{thebibliography}{99}
\bibitem{mech}M. Chaichian, M.M. Sheikh-Jabbari, A. Tureanu, Phys.
Rev. Lett. 86, 2716 (2001)
\bibitem{mechnext}A. Deriglazov, JHEP 0303, 021 (2003)
\bibitem{qmnext} S. Ghosh, Phys. Lett. B 648, 262 (2007)
\bibitem{field}M. Chaichian, P. Pre\v{s}najder and  A. Tureanu,
Phys. Rev. Lett. 94, 151602 (2005)
\bibitem{snyder}H.S. Snyder, Phys. Rev. 72, 68 (1947)
\bibitem{grav1}S. Doplicher, K. Fredenhagen, J.E. Roberts, Phys. Lett. B 331, 39
(1994)
\bibitem{string1}A. Connes, M.R. Douglas, A. Schwarz, JHEP 9802, 003
(1998)
\bibitem{1a}S. Coleman, S.L. Glashow, Phys. Rev. D 59, 116008
(1999)
\bibitem{1anext}
R.J. Protheore, H. Meyer, Phys. Lett. B 493, 1 (2000)
\bibitem{class1}S. Zakrzewski
; q-alg/9602001
\bibitem{class2}
Y. Brihaye, E. Kowalczyk, P. Maslanka
; math/0006167
\bibitem{conservative}M. Daszkiewicz, Acta Phys. Pol. B (in press)
\bibitem{oeckl}R. Oeckl, J. Math. Phys. 40, 3588 (1999)
\bibitem{chi}M. Chaichian, P.P. Kulish, K. Nashijima, A. Tureanu, Phys. Lett. B
604, 98 (2004)
\bibitem{dasz1}M. Daszkiewicz,
Mod. Phys. Lett. A 23, 505 (2008)
\bibitem{kappaP}J. Lukierski, A. Nowicki, H. Ruegg and V.N. Tolstoy, Phys. Lett.
B 264, 331 (1991)
\bibitem{kappaG}S. Giller, P. Kosinski, M. Majewski, P. Maslanka
and J. Kunz, Phys. Lett. B 286, 57 (1992)
\bibitem{lie1}
J. Lukierski and M. Woronowicz, Phys. Lett. B 633, 116 (2006)
\bibitem{qdef}O. Ogievetsky, W.B.  Schmidke, J. Wess, B. Zumino, Comm. Math. Phys.
150, 495 (1992)
\bibitem{paolo}
P. Aschieri, L. Castellani, A.M. Scarfone, Eur. Phys. J. C 7, 159
(1999)
\bibitem{nh}M. Daszkiewicz, Acta Phys. Pol. B 41, 1889 (2010)
\bibitem{nh1}M. Daszkiewicz, Mod. Phys. Lett. A 24, 1325 (2009)
\bibitem{bloch}C. Blohmann, J. Math. Phys. 44, 4736 (2003)
\bibitem{wess}J. Wess
; hep-th/0408080
\bibitem{sch}E. Schr\"{o}dinger, Naturwissenschaften 14, 664 (1926)
\bibitem{glaubner}R.J. Glauber, Phys. Rev. 131, 2766 (1963)
\bibitem{ligo}LIGO; http://www.ligo.caltech.edu/
\bibitem{23} K. Bolonek, P. Kosi\'nski, Phys. Lett. B 547, 51 (2002)
\bibitem{23a}K. Bolonek, P. Kosi\'nski, Acta Phys. Polon. B 34, 2575 (2003)
\bibitem{24} L. J. Schiff, {\em Quantum Mechanics}\/, McGraw-Hill, New York 1968
\bibitem{24a}S. Kryszewski, {\em Quantum Mechanics}\/, University of Gdansk, Gdansk 2010
\end{thebibliography}
\end{document}